\documentclass[conference]{IEEEtran}
\IEEEoverridecommandlockouts
\usepackage{cite}
\usepackage{amsmath,amssymb,amsfonts}
\usepackage{algorithmic}
\usepackage{graphicx}
\usepackage{textcomp}
\usepackage{xcolor}
\usepackage{bm}
\usepackage{multirow}
\usepackage{tikz}
\usepackage{pgfplots}
\usepackage{mathtools}
\usepackage[ruled,vlined]{algorithm2e}
\def\BibTeX{{\rm B\kern-.05em{\sc i\kern-.025em b}\kern-.08em
    T\kern-.1667em\lower.7ex\hbox{E}\kern-.125emX}}

\DeclareMathOperator{\len}{\ell}
\DeclarePairedDelimiter\ceil{\lceil}{\rceil}

\begin{document}

\title{Compressing Piecewise Smooth Images\\
	   with the Mumford-Shah Cartoon Model\\
\thanks{This work has received funding from the European Research Council
	(ERC) under the European Union's Horizon 2020 research and
	innovation programme (grant agreement no. 741215, ERC Advanced
	Grant INCOVID).}
}

% \author{
%   \IEEEauthorblockN{Ferdinand Jost, Pascal Peter, and Joachim Weickert}
%   \IEEEauthorblockA{Mathematical Image Analysis Group, 
%       Faculty of Mathematics and Computer Science\\
%       Campus E1.7, Saarland University, 66041 Saarbr\"ucken, Germany\\
%       \{jost, peter, weickert\}@mia.uni-saarland.de}

\author{\IEEEauthorblockN{Ferdinand Jost}
  \IEEEauthorblockA{Mathematical Image Analysis Group\\
   Saarland Informatics Campus\\
   Saarland University, Campus E1.7\\
   66041 Saarbr{\"u}cken, Germany\\
   jost@mia.uni-saarland.de}
\and
\IEEEauthorblockN{Pascal Peter}
  \IEEEauthorblockA{Mathematical Image Analysis Group\\
   Saarland Informatics Campus\\
   Saarland University, Campus E1.7\\
   66041 Saarbr{\"u}cken, Germany\\
   peter@mia.uni-saarland.de}
\and
\IEEEauthorblockN{Joachim Weickert}
  \IEEEauthorblockA{Mathematical Image Analysis Group\\
   Saarland Informatics Campus\\
   Saarland University, Campus E1.7\\
   66041 Saarbr{\"u}cken, Germany\\
   weickert@mia.uni-saarland.de}
}

\maketitle

\begin{abstract}
Compressing piecewise smooth images is important for many data 
types such as depth maps in 3D videos or optic flow fields for motion 
compensation. Specialised codecs that rely on explicitly stored 
segmentations excel in this task since they preserve discontinuities 
between smooth regions. However, current approaches rely on ad hoc 
segmentations that lack a clean interpretation in terms of energy
minimisation. As a remedy, we derive a generic region merging 
algorithm from the Mumford-Shah cartoon model. It adapts the 
segmentation to arbitrary reconstruction operators for the segment 
content. In spite of its conceptual simplicity, our framework can 
outperform previous segment-based compression methods as well as 
BPG by up to 3~dB.
\end{abstract}


%%%%%%%%%%%%%%%%%%%%%%%%%%%%%%%%%%%%%%%%%%%%%%%%%%%%%%%%%%%%%%%%%%%%%%%%%

\section{Introduction}

Piecewise smooth data form an important sub-class of visual content. 
This includes not only cartoon-like images, but also depth maps in 
3D videos, and optic flow fields that represent interframe motion. 
These images have in common that edges are their most salient features. 
Consequentially, specialised codecs for this image type should model 
discontinuities with high accuracy. 

\medskip
Mainberger et al. \cite{MBWF11} have shown that inpainting-based
compression is an excellent match for cartoon-like images:
They extract and store image edges with sampled pixel data and
interpolate the segment interior with homogeneous diffusion.
This work acted as a blueprint for a series of depth map compression
approaches \cite{GLG12,LSJO12,HMWP13} that rely on the same basic
principle. All these methods use heuristic segmentation methods 
that do not rely on optimality concepts such as energy minimisation. 
Since segmentation methods have benefitted a lot from energy 
minimisation concepts, the natural question arises if this is also 
useful for compressing piecewise smooth data.

%-------------------------------------------------------------------------

\subsection{Our Contribution}

The goal of our paper is to give an answer to this question. To
this end, we consider the simplest and best understood energy-based
segmentation method: the Mumford--Shah cartoon model. It consists
of a data term that measures the approximation quality and a regulariser
which penalises the length of the segmentation boundaries. 
The original Mumford-Shah cartoon model approximates the data within
each segment by their average grey value. For most compression approximations
this is too crude. As a remedy, we use an inpainting-based approximation
which relies on data specified on a sparse regular grid. Our 
experiments will show that this adaptation can lead to codecs that
are simple, fast, and qualitatively significantly superior to a 
state-of-the-art method in segment-based compression~\cite{HMWP13},
as well as the transform-based Better Portable Graphics (BPG)~\cite{Be14}. 
Moreover, our framework is fairly generic: Within each segment, we
can use arbitrary reconstructions by inpainting as well as
reconstructions by polynomial approximations.

%------------------------------------------------------------------------

\subsection{Related Work}

Generic codecs designed for natural images can also be used for 
compressing piecewise smooth images. For instance, Better Portable 
Graphics (BPG)~\cite{Be14}, a container format for the intra coding in 
HEVC~\cite{SOHW12}, marks a state-of-the-art method for transform coding. 
Here, coarsely quantised coefficients of a cosine transform represent 
the image. 

\medskip
Dedicated codecs for piece-wise smooth data consider the geometric 
information of the discontinuities in some way. Transform-based video 
plus depth codecs such as HEVC-3D also contain depth map specific 
coding techniques \cite{TYMOVW15}. For instance, Merkle~et~al.~\cite{MMMW16} 
improve performance on depth maps by introducing geometric information 
to block-based prediction steps. Georgiev et al. \cite{GBG15} 
explicitly use image segmentation to downsample depth maps for 
efficient storage in video applications. 

\medskip
As an alternative to inpainting-based recontructions within each
segment, also constant or planar approximations have been advocated
\cite{OA14,ST13,ST14,Jae11}. Our framework can accomodate these
approaches as well, since the reconstruction method is handled as
a black box process. In our experiments, we will also compare with
these approximation methods.

%------------------------------------------------------------------------

\subsection{Paper Structure}

We introduce our new framework in Section \ref{sec:framework} and 
derive example codecs from it in Section \ref{sec:examples}. 
After an experimental evaluation in Section \ref{sec:experiments} 
we conclude our paper with a discussion and an outlook on future 
work in Section \ref{sec:conclusion}.

%%%%%%%%%%%%%%%%%%%%%%%%%%%%%%%%%%%%%%%%%%%%%%%%%%%%%%%%%%%%%%%%%%%%%%%%%%

\section{Compression Framework}\label{sec:framework}

In the following, we generalise the Mumford-Shah cartoon model~\cite{MS89,KLM94}
and use it as the foundation of our framework for segment-based compression. 
For its minimisation, we adapt a straightforward region merging algorithm. 

%-------------------------------------------------------------------------

\subsection{The Generalised Mumford-Shah Cartoon Model}

For a given image $f(\bm{x}) : \Omega \rightarrow \mathbb{R}$ with 
a rectangular image domain $\Omega \subset \mathbb{R}^2$ we want 
to partition $\Omega$ into segments $\Omega_i$. 
The reconstruction $u$ inside each of these segments should be 
as close as possible to $f$ and the set of boundaries $K$ should 
be inexpensive to store. We express this optimisation problem with 
the following generalised Mumford-Shah functional:
\begin{equation}\label{eq:mumfordshah}
  E(u, K) = \int_{\Omega \setminus K} (u(\bm{x}) - f(\bm{x}))^2 \, 
  d\bm{x} + \lambda \len(K) \, .
\end{equation}
Here, $\len(K)$ specifies the length of segment boundaries, and 
$\lambda$ is a scalar rate-distortion weight. The first part of 
this functional is a data term that minimises the reconstruction 
error of our compression method. The second part can be seen as a 
cost term that approximates the coding cost of the segment boundaries.

\medskip
The classical Mumford-Shah cartoon model assumes $u$ to be a 
piecewise constant approximation of the original image~$f$. 
In contrast, our framework treats the reconstruction $u$ as 
a black box model, allowing arbitrary operators.

%-------------------------------------------------------------------------

\subsection{Minimisation Strategy}

We minimise the energy functional from Eq.~(\ref{eq:mumfordshah}) with
the region merging algorithm of Koepfler~et~al.~\cite{KLM94}. 
It merges adjacent regions $\Omega_i$ and $\Omega_j$ if this 
decreases the energy, i.e. if
\begin{equation}\label{eq:2normal}
	E(u,K) - E(u', K \setminus \partial (\Omega_i, \Omega_j)) > 0 \, .
\end{equation}
Here $\partial (\Omega_i, \Omega_j)$ denotes the joint boundary 
between segments $\Omega_i$ and $\Omega_j$, $u$ is the reconstruction 
before merging, and $u'$ is the reconstruction after merging the regions. 

\medskip
Similar to Koepfler~et~al.~\cite{KLM94} for piece-wise constant 
approximations, we simplify this criterion in our general case. 
As regions are independent of each other, $u$ and $u'$ only differ 
in the merged region $\Omega_i \cup \Omega_j$. This allows us to 
localise the merging criterion (\ref{eq:2normal}) to this area and get
\begin{equation}\label{eq:mergingcriterion}
 \frac{\int_{\Omega_i \cup \Omega_j}
       (u'(\bm{x}) - f(\bm{x}))^2 - (u(\bm{x}) - f(\bm{x}))^2 \, d\bm{x}}
      {\len(\partial (\Omega_i, \Omega_j))}
 < \lambda \, .
\end{equation}
The numerator of this fraction gives us the increase of the approximation 
error, whereas the denominator can be seen as the coding cost reduction
of the segment boundaries. We can steer this tradoff by means of the 
weighting parameter $\lambda$.

\medskip
With this merging criterion we only have to determine an order in 
which we merge our candidates. Similar to the original approach of 
Koepfler~et~al. and the GSO algorithm of Schiopu and Tabus~\cite{ST13} 
we use a greedy strategy. We always choose the pair of regions 
that decreases our energy the most, i.e. requires the smallest value 
for $\lambda$ to be able to merge in Eq.~\eqref{eq:mergingcriterion}. 
This yields the following region merging algorithm:

%...........................................................................

\begin{algorithm}\label{alg:2normal}
 \SetAlgoLined
 \SetArgSty{textrm}
 \textbf{Input:} original image $f:\Omega \rightarrow \mathbb{R}$, 
                 weight $\lambda$. \\
 \textbf{Output:} segmentation $(u, K)$.\\
	
 Initialise segmentation as one region for each pixel.\\
	
 \ForAll{neighbouring regions $(\Omega_i, \Omega_j)$}
        {
	Compute minimal $\lambda_{i, j}$ allowing $\Omega_i$, $\Omega_j$ 
        to merge.
	}
 \While{minimal $\lambda_{i, j} < \lambda$}
       {
       Merge regions $\Omega_i$, $\Omega_j$ with minimal $\lambda_{i, j}$ \\ 
       Update $\lambda_{i, j}$ for merged region and its neighbours.
       }
 \Return current segmentation
\caption{Proposed region merging}
\end{algorithm}

%-------------------------------------------------------------------------

\subsection{Encoding}

For a fixed reconstruction method, the proposed optimisation strategy 
in Algorithm \ref{alg:2normal} gives us both a locally optimal 
segmentation, as well as the corresponding reconstruction $u$ in 
each region. Consequentially, an encoder needs to store the segmentation 
itself and all necessary data to restore the approximation inside segments. 

\medskip
We store the segment boundaries with chain codes similar to 
Hoffmann~et~al.~\cite{HMWP13}. They describe the boundaries in 
terms of a list of starting elements and a sequence of directions. 
Since segment boundaries lie between pixels, there are only three 
possible directions to follow, which makes this approach very efficient. 
It also reflects our cost model from Eq.~\eqref{eq:mumfordshah} where 
we approximate the coding cost of segment boundaries by their length.

\medskip
The data we have to store for the approximation inside segments depend 
on the reconstruction operator: For a polynomial approximation this 
can be a list of coefficients for each segment, while inpainting-based 
approaches require sparse pixel data inside the segment. 
Depending on the nature of these data we can further reduce the
coding cost using e.g. quantisation techniques. More details about
concrete codecs can be found in Section \ref{sec:examples}.

\medskip
The final payload, consisting of the segmentation in form of a 
chain code and the data required to restore the approximations 
inside segments, is then passed to an entropy coder. 
In our case we use \emph{lpaq2}~\cite{Ma05}. 

%-------------------------------------------------------------------------

\subsection{Decoding}

The decoding step in our framework is straightforward: We first use 
\emph{lpaq2} to recover the chain code and the payload for the 
approximations inside segments. Afterwards we construct our 
segmentation by following the chain codes starting at the stored 
reference points. Finally, we individually compute the approximation 
inside each segment to get the decoded image.

%%%%%%%%%%%%%%%%%%%%%%%%%%%%%%%%%%%%%%%%%%%%%%%%%%%%%%%%%%%%%%%%%%%%%%%%%%

\section{Example Codecs}\label{sec:examples}

Let us now discuss concrete codecs for two classes of popular 
reconstruction operators: inpainting-based reconstructions and
approximation methods based on two-dimensional polynomials.

%-------------------------------------------------------------------------

\subsection{Inpainting-based Codecs}

Inpainting approaches for compression use only a small 
fraction of known data to interpolate the missing values inbetween. 
For each segment $\Omega_i$ we assume that the values are known
in a subset $K_i \subset \Omega_i$, which is called the
inpainting mask. Reconstructing the image in $\Omega_i \setminus K_i$ 
comes down to solving an inpainting problem 
\begin{equation}
  L u(\bm{x}) = 0 \qquad \text{for}\quad \bm{x} 
  \in \Omega_i \setminus K_i
\end{equation}
with fixed mask points and reflecting boundary conditions across 
segment boundaries:
\begin{align}
 u(\bm{x}) = f(\bm{x}) &\qquad\text{for}\quad \bm{x} \in K_i, \\
 \partial_{\bm{n}} u(\bm{x}) = 0 &\qquad\text{for}\quad \bm{x} 
   \in \partial \Omega_i\,.
\end{align}
Here $L$ is a suitable inpainting operator, and $\bm{n}$ is the 
normal vector to the segment boundary $\partial \Omega_i$.
For $L$, we choose the Laplacian $\Delta$, resulting in inpainting 
with homogeneous diffusion~\cite{Ii62}. This operator 
is used in many segment-based codecs~\cite{HMWP13,MBWF11,GLG12}. 
We discretise the continuous inpainting problem with standard finite
differences and solve the resulting linear system of equations
iteratively with a conjugate gradient algorithm \cite{MM05}.

\medskip
We also use Shepard interpolation~\cite{Sh68} as an inpainting operator
within each segment. 
In the discrete setting, let $\bm{x}_k$ denote the location of pixel
$k$, and let $f_k:=f(\bm{x}_k)$. Furthermore, let $M_i$ be the index set
of the mask pixels of the $i$-th segment. 
Then Shepard interpolation reconstructs the value $u_j$ in a
non-mask pixel $j$ of this segment as a weighted average of the 
values in the mask pixels:
\begin{equation}\label{eq:shepard}
  u_j = \frac{\sum_{k \in M_i} w(\bm{x}_k \!-\! \bm{x}_j) \, f_k}
             {\sum_{k \in M_i} w(\bm{x}_k \!-\! \bm{x}_j)} \, .
\end{equation}
A popular choice for the weighting function $w$ is a Gaussian with 
standard deviation $\sigma$, truncated outside 
$(\ceil*{4 \sigma} + 1) \times (\ceil*{4 \sigma} + 1)$. 
Following \cite{AAS17}, we adapt $\sigma$ to the mask density $d$ 
(ratio of mask pixels and image pixels) by 
choosing $\sigma := 1/\sqrt{\pi d}$. 
Due to the localisation to this window, Shepard interpolation can be 
substantially faster than homogeneous diffusion inpainting~\cite{AAS17,Pe19}.

\medskip
Inpainting requires to store the mask location along with their
corresponding values of the original image $f$. To reduce the 
overhead from storing 
pixel positions, we select mask pixels on a regular grid that can 
be reconstructed by only storing its density parameter $d$. 
A further reduction of the coding cost can be achieved by 
a coarser quantisation of the grey values.

\medskip
A common strategy in inpainting-based compression is tonal optimisation: 
Instead of storing the original pixel values, we allow deviations 
to other quantisation levels if this yields an overall lower 
reconstruction error. For homogeneous diffusion we achieve this 
with the strategy of Mainberger~et~al.~\cite{MHWT12}, and for Shepard 
interpolation with the approach of Peter~\cite{Pe19}.  

%-------------------------------------------------------------------------

\subsection{Approximation-based Codecs}

An alternative approach to obtain a reconstruction inside a segment 
$\Omega_i$ is to approximate the image with a function from the 
class $P_n$ of bivariate polynomials of degree $n$. The 
goal is to find the polynomial $p_n(\bm{x})$ that minimises the 
quadratic error w.r.t. the original image $f$:
\begin{equation}
\label{eq:least_squares}
 \min_{p_n \in P_n} \left\lVert f(\bm{x}) - p_n(\bm{x})
 \right\rVert_2^2 
 \qquad\text{for}\quad \bm{x} \in \Omega_i \,.
\end{equation}
In the case of $n = 0$ and $n = 1$ this covers constant and planar 
reconstructions~\cite{ST13,ST14,Jae11,OA14}.

\medskip
Eq.~\eqref{eq:least_squares} is a classical least squares
problem, leading to a system of linear equations \cite{Sch89}. Its 
unknowns are the coefficients of the polynomial. Tonal optimisation 
is not necessary, since these coefficients already give the best 
approximation.

\medskip
To specify the bivariate polynomial $p_n(\bm{x})$ we have to store
a list of ${n + 2 \choose n}$ coefficients.
As it is important to preserve the precision of those coefficients, 
we store them as $32$-bit floating-point values.

%%%%%%%%%%%%%%%%%%%%%%%%%%%%%%%%%%%%%%%%%%%%%%%%%%%%%%%%%%%%%%%%%%%%%%%%%%%

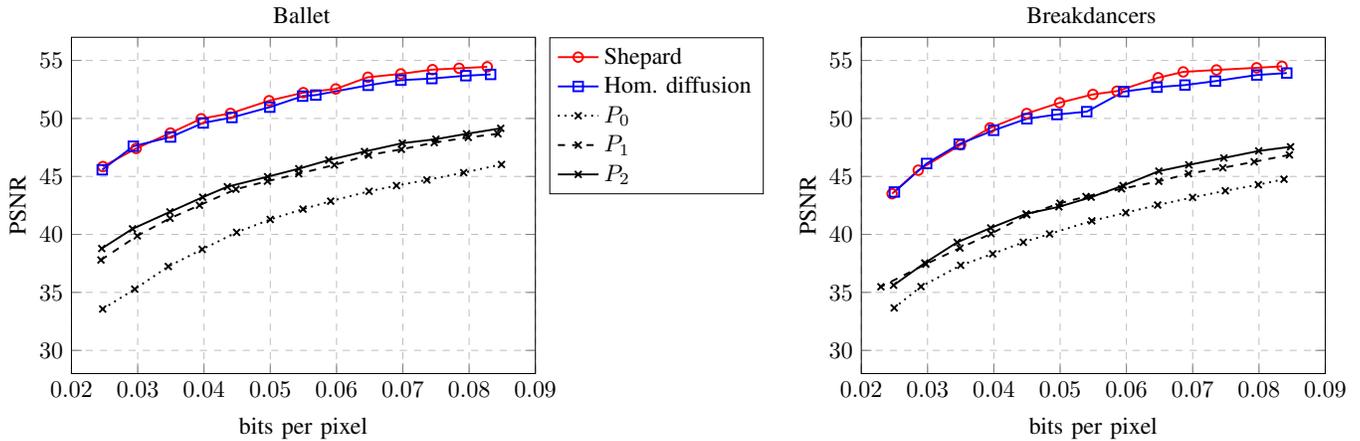
\begin{figure*}[ht]
  \centering
  \begin{tikzpicture}[scale=0.9]
    \pgfplotsset{every mark/.append style={solid}}
	\begin{axis}[
	  label={test},
	  y=0.60cm/3.5,
	  xlabel={bits per pixel},
	  ylabel={PSNR},
	  y label style={at={(axis description cs:0.1,.5)},anchor=south},
	  xtick={0.02, 0.03, 0.04, 0.05, 0.06, 0.07, 0.08, 0.09},
	  ytick={30,35,40,45,50,55},
	  xmin=0.02, xmax=0.09,
	  ymin=28, ymax=57,
	  legend pos=outer north east,
	  ymajorgrids=true,
	  xmajorgrids=true,
	  grid style=dashed,
	  legend cell align={left},
	  scaled ticks=false, tick label style={/pgf/number format/fixed},
	]
	\addplot[color=red, mark=o, style=thick, mark options=solid] 
	  table[x index=4,y index=5]{resources/ballet_compare.txt};
	\addplot[color=blue, mark=square, style=thick, mark options=solid] 
	  table[x index=8,y index=9]{resources/ballet_compare.txt};
	\addplot[color=black, mark=x, style=thick, dotted, mark options=solid] 
	  table[x index=10,y index=11]{resources/ballet_compare.txt};
	\addplot[color=black, mark=x, style=thick, dashed, mark options=solid] 
	  table[x index=12,y index=13]{resources/ballet_compare.txt};
	\addplot[color=black, mark=x, style=thick, mark options=solid] 
	  table[x index=14,y index=15]{resources/ballet_compare.txt};
	  
	\legend{Shepard\\Hom. diffusion\\$P_0$\\$P_1$\\$P_2$\\}
	\end{axis}
	
	\node at (3.4,5.3) {\small Ballet};
  \end{tikzpicture}
~
  \begin{tikzpicture}[scale=0.9]
	\pgfplotsset{every mark/.append style={solid}}
	\begin{axis}[
	  y=0.60cm/3.5,
	  xlabel={bits per pixel},
	  ylabel={PSNR},
	  y label style={at={(axis description cs:0.1,.5)},anchor=south},
	  xtick={0.02, 0.03, 0.04, 0.05, 0.06, 0.07, 0.08, 0.09},
	  ytick={30,35,40,45,50,55},
	  xmin=0.02, xmax=0.09,
	  ymin=28, ymax=57,
	  ymajorgrids=true,
	  xmajorgrids=true,
	  grid style=dashed,
	  legend cell align={left},
	  scaled ticks=false, tick label style={/pgf/number format/fixed},
    ]
    \addplot[color=red, mark=o, style=thick, mark options=solid] 
      table[x index=4,y index=5]{resources/breakdancers_compare.txt};
    \addplot[color=blue, mark=square, style=thick, mark options=solid] 
      table[x index=8,y index=9]{resources/breakdancers_compare.txt};
    \addplot[color=black, mark=x, style=thick, dotted, mark options=solid] 
      table[x index=10,y index=11]{resources/breakdancers_compare.txt};
    \addplot[color=black, mark=x, style=thick, dashed, mark options=solid] 
      table[x index=12,y index=13]{resources/breakdancers_compare.txt};
    \addplot[color=black, mark=x, style=thick, mark options=solid] 
      table[x index=14,y index=15]{resources/breakdancers_compare.txt};
    \legend{}
  \end{axis}
	
  \node at (3.4,5.3) {\small Breakdancers};
  \end{tikzpicture}
  \caption{Comparison of different reconstruction operators within 
           our framework. $P_0$--$P_2$ denote polynomial approximations
           of degree $0$ to $2$. 
           Inpainting approaches clearly outperform the approximation methods.}
	\label{fig:graph_operators}
\end{figure*}

\begin{figure*}[ht]
  \centering
  \setlength{\tabcolsep}{4pt}
  \begin{tabular}{c c c c c}
    ~ & Ballet & Ballet zoomed & Breakdancers & Breakdancers zoomed\\
        
    \multirow{1}{*}[12ex]{\rotatebox[origin=c]{90}{Original}} &
    \includegraphics[width=0.22\textwidth]{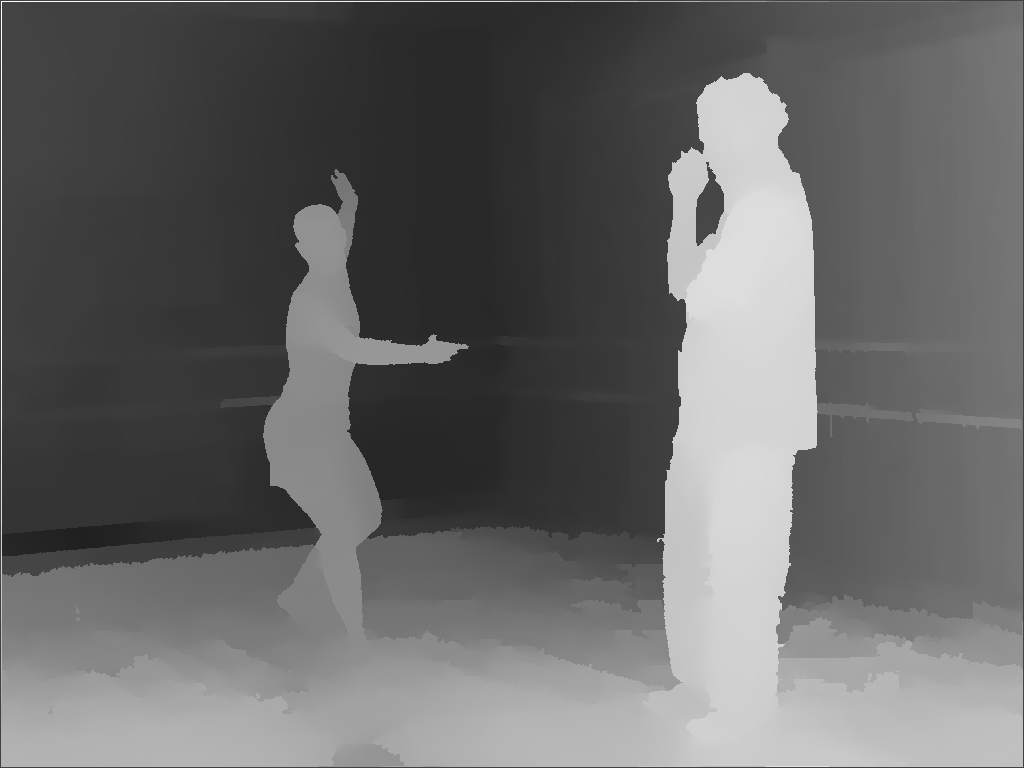}&
    \includegraphics[trim=766 69 105 584,clip=true,width=0.22\textwidth]
                    {resources/ballet.png}&
    \includegraphics[width=0.22\textwidth]
                    {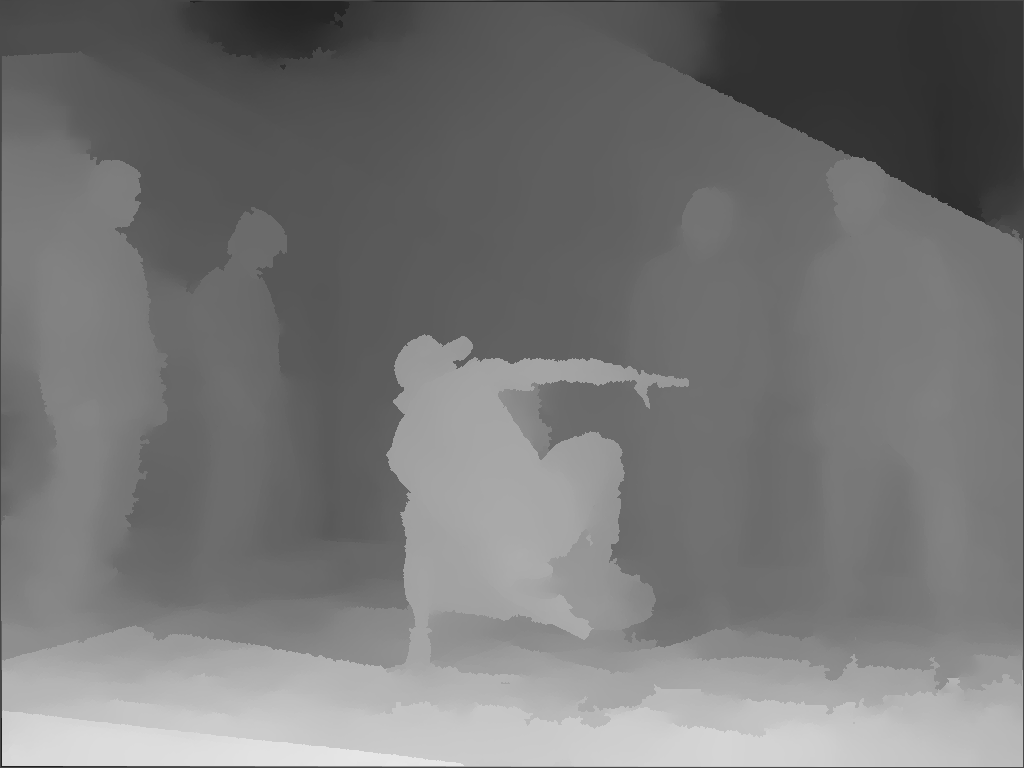}&
    \includegraphics[trim=473 300 420 370,clip=true,width=0.22\textwidth]
                    {resources/breakdancers.png}\\
        
    \multirow{1}{*}[12ex]{\rotatebox[origin=c]{90}{\color{red}\bf{Shepard}}} &
    \includegraphics[width=0.22\textwidth]{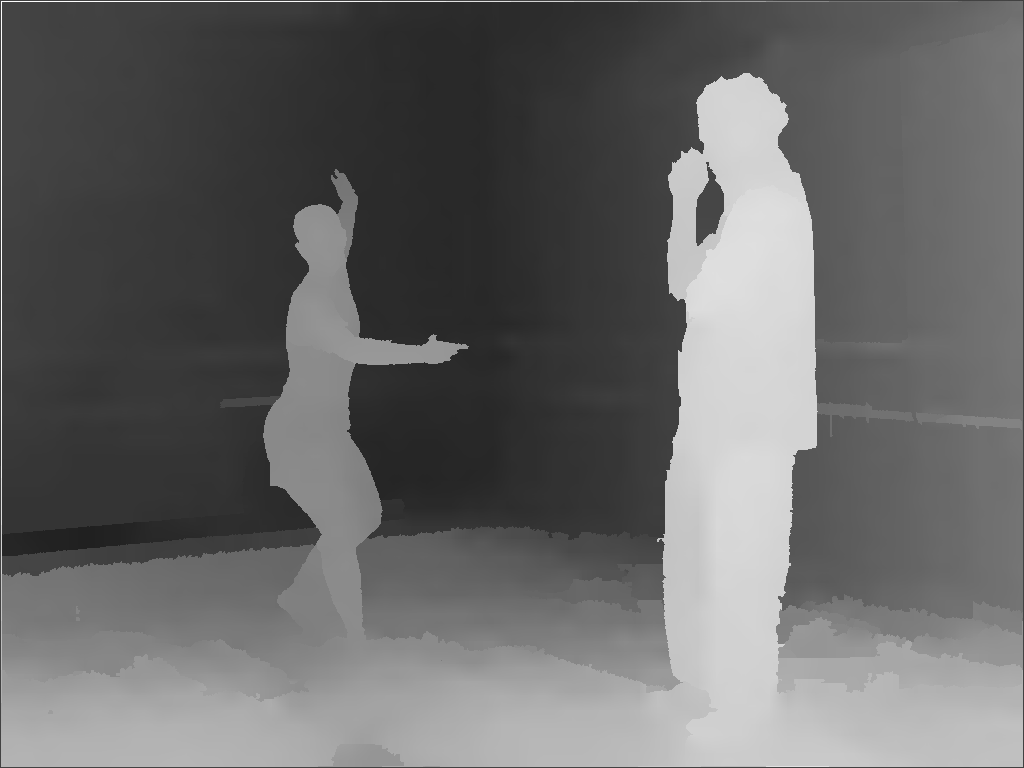}&
    \includegraphics[trim=766 69 105 584,clip=true,width=0.22\textwidth]
                    {resources/ballet_shp_0-04.png}&
    \includegraphics[width=0.22\textwidth]
                    {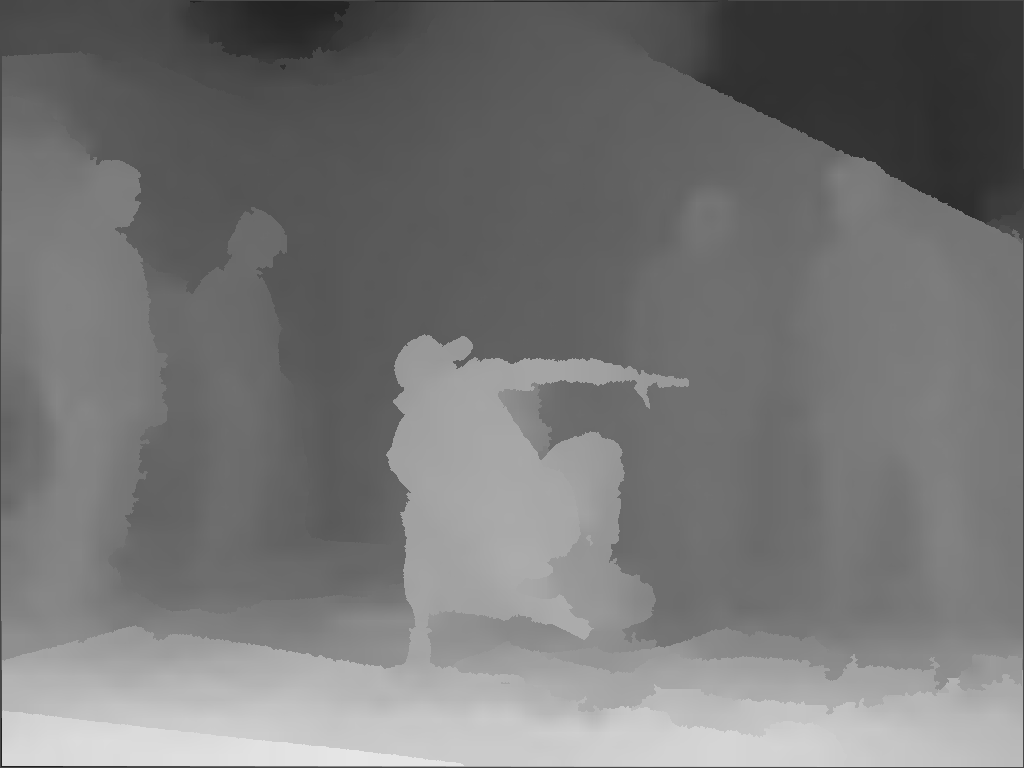}&
    \includegraphics[trim=473 300 420 370,clip=true,width=0.22\textwidth]
                    {resources/breakdancers_shp_0-04.png}\\
        
    \multirow{1}{*}[10ex]{\rotatebox[origin=c]{90}{BPG}} &
    \includegraphics[width=0.22\textwidth]
                    {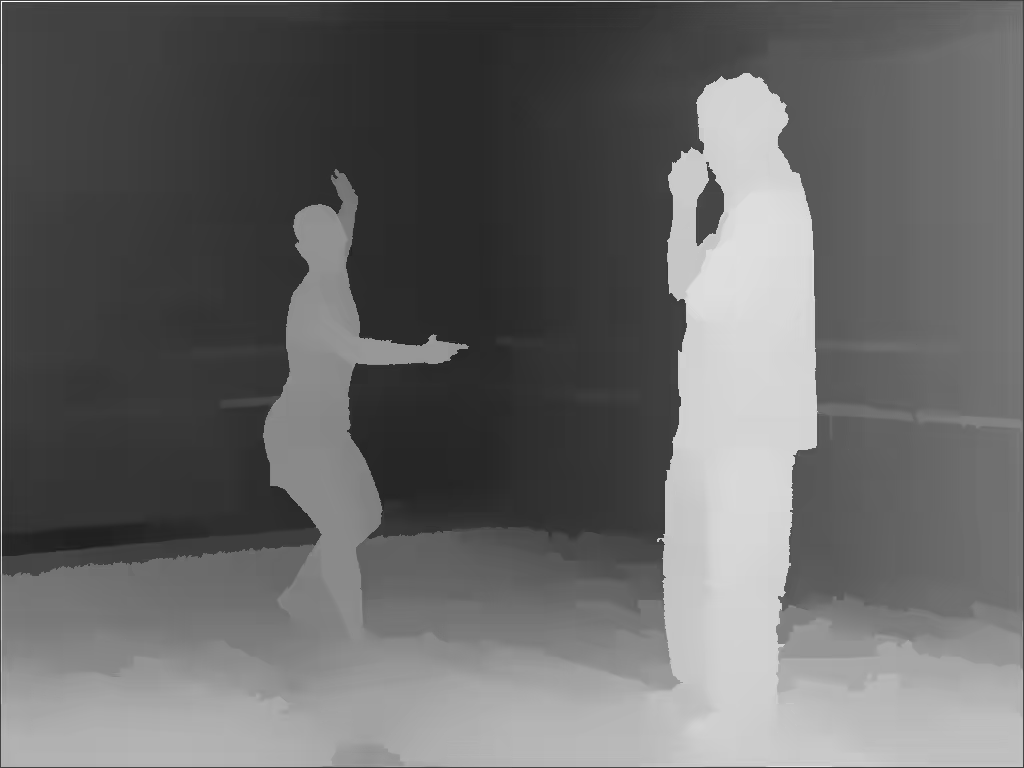}&
    \includegraphics[trim=766 69 105 584,clip=true,width=0.22\textwidth]
                    {resources/ballet_bpg_0-04.png}&
    \includegraphics[width=0.22\textwidth]
                    {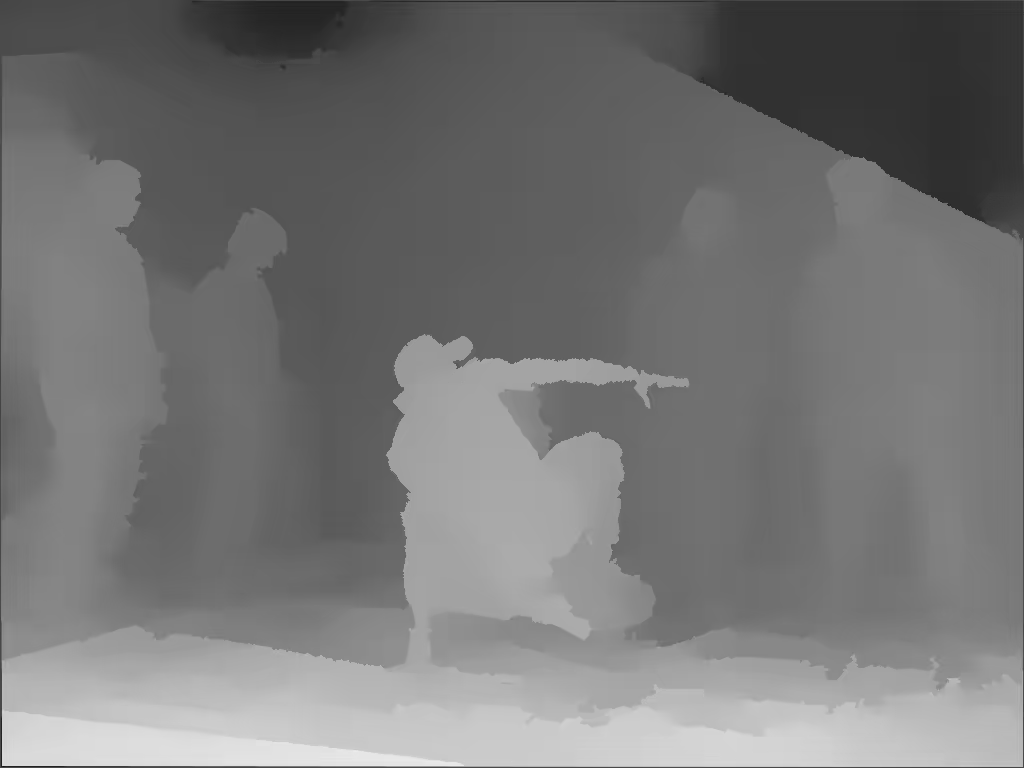}&
    \includegraphics[trim=473 300 420 370,clip=true,width=0.22\textwidth]
                    {resources/breakdancers_bpg_0-04.png}\\
  \end{tabular}
    
  \caption{Compression results of different methods for a compression
           rate of 0.04 bpp. Our framework offers sharper and more precise 
           edges than BPG.}
  \label{fig:images}
\end{figure*}

%............................................................................

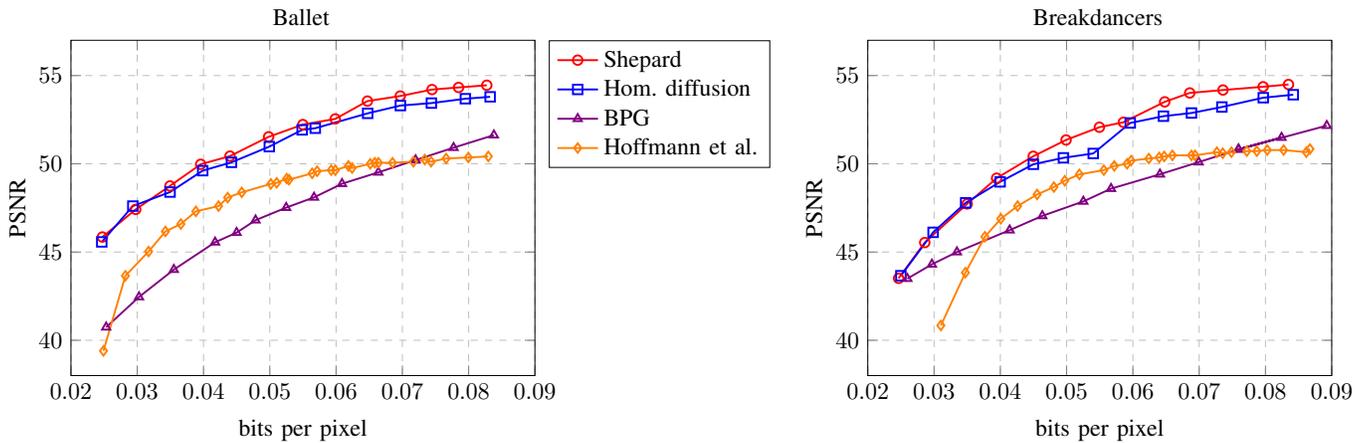
\begin{figure*}[ht]
  \centering
  \begin{tikzpicture}[scale=0.9]
  \pgfplotsset{every mark/.append style={solid}}
  \begin{axis}[
    y=0.60cm/2.3,
    xlabel={bits per pixel},
	ylabel={PSNR},
	y label style={at={(axis description cs:0.1,.5)},anchor=south},
	xtick={0.02, 0.03, 0.04, 0.05, 0.06, 0.07, 0.08, 0.09},
	ytick={40,45,50,55},
	xmin=0.02, xmax=0.09,
	ymin=38, ymax=57,
	legend pos=outer north east,
	ymajorgrids=true,
	xmajorgrids=true,
	grid style=dashed,
	legend cell align={left},
	scaled ticks=false, tick label style={/pgf/number format/fixed},
  ]
  \addplot[color=red, mark=o, style=thick, mark options=solid] 
	table[x index=4,y index=5]{resources/ballet_compare.txt};
  \addplot[color=blue, mark=square, style=thick, mark options=solid] 
	table[x index=8,y index=9]{resources/ballet_compare.txt};
  \addplot[color=violet, mark=triangle, style=thick, mark options=solid] 
	table[x index=0,y index=1]{resources/ballet_compare.txt};
  \addplot[color=orange, mark=diamond, style=thick, mark options=solid] 
	table[x index=0,y index=1]{resources/hoffmann_ballet.txt};
  \legend{Shepard\\Hom. diffusion\\BPG\\Hoffmann et al.\\}
  \end{axis}
	
  \node at (3.4,5.3) {\small Ballet};
  \end{tikzpicture}
	~
  \begin{tikzpicture}[scale=0.9]
	\pgfplotsset{every mark/.append style={solid}}
	\begin{axis}[
	  y=0.60cm/2.3,
	  xlabel={bits per pixel},
	  ylabel={PSNR},
	  y label style={at={(axis description cs:0.1,.5)},anchor=south},
	  xtick={0.02, 0.03, 0.04, 0.05, 0.06, 0.07, 0.08, 0.09},
	  ytick={40,45,50,55},
	  xmin=0.02, xmax=0.09,
	  ymin=38, ymax=57,
	  ymajorgrids=true,
	  xmajorgrids=true,
	  grid style=dashed,
	  legend cell align={left},
	  scaled ticks=false, tick label style={/pgf/number format/fixed},
	]
	\addplot[color=red, mark=o, style=thick, mark options=solid] 
	  table[x index=4,y index=5]{resources/breakdancers_compare.txt};
	\addplot[color=blue, mark=square, style=thick, mark options=solid] 
	  table[x index=8,y index=9]{resources/breakdancers_compare.txt};
	\addplot[color=violet, mark=triangle, style=thick, mark options=solid] 
	  table[x index=0,y index=1]{resources/breakdancers_compare.txt};
	\addplot[color=orange, mark=diamond, style=thick, mark options=solid] 
	  table[x index=0,y index=1]{resources/hoffmann_breakdancers.txt};
	\legend{}
	\end{axis}
	
	\node at (3.4,5.3) {\small Breakdancers};
	\end{tikzpicture}
	\caption{Comparison of different reconstruction operators to BPG 
                 and Hoffmann~et~al.~\cite{HMWP13}. In both images, 
                 Hoffmann~et~al.~ outperforms BPG for medium compression 
                 ratios while our framework beats both competitors 
                 consistently over the full range of bitrates.}
	\label{fig:graph_comparison}
\end{figure*}

%............................................................................

\section{Experiments}\label{sec:experiments}

Let us now present results of our framework for five different 
reconstruction operators: inpainting with homogeneous diffusion, 
Shepard interpolation, and polynomial approximations of degrees 
zero to two. Furthermore, we compare against BPG~\cite{Be14} 
and the state-of-the-art segment-based codec of Hoffmann~et~al.~\cite{HMWP13}. 
To this end we use the depth maps \emph{ballet} and \emph{breakdancers} 
from the MVD sequence~\cite{Zi04}. We optimise the weight $\lambda$ as well
as the mask density and number of quantisation levels of our inpainting-based
approaches with a grid search.

\medskip
Comparing the five operators in Fig.~\ref{fig:graph_operators} within 
our framework reveals a clear superiority of inpainting over approximation. 
Shepard interpolation and homogeneous diffusion outperform the best 
polynomial approach by more than 5~dB in the 
peak-signal-to-noise-ratio (PSNR). There are two reasons for this 
large superiority: On one hand, the inpainting operators require 
less segments to achieve good overall reconstructions. 
%(see Table~\ref{tab:number_segments}). 
On the other hand, approximation approaches suffer from the overhead 
of high precision coefficients, compared with the quantised grey values 
stored by the inpainting methods. Overall, Shepard interpolation performs    
best: It does not only offer a slight qualitative advantage over 
homogeneous diffusion, but is also faster by almost a factor 10. 

\medskip
Therefore, we also choose Shepard interpolation for our comparison with 
existing approaches. Fig.~\ref{fig:images} visualises the results.
For both test images we observe that our framework yields considerably 
sharper and more faithful edges than BPG. 

\medskip
This is also reflected in the quantitative analysis that is presented
in the rate-distortion curves in Fig.~\ref{fig:graph_comparison}.
The codec of Hoffmann~et~al.~\cite{HMWP13} beats BPG for medium
compression ratios, but remains behind for high and low ratios. 
In comparison, our framework outperforms both competitors for the 
full range of bitrates, often by a margin of more than 3~dB. 
Moreover, by employing Shepard inpainting, it also uses a faster 
reconstruction that the codec of Hoffmann~et~al., which relies 
on homogeneous diffusion inpainting. 
Thus, our approach offers clear advantages, both qualitatively and 
in terms of algorithmic efficiency. 

%%%%%%%%%%%%%%%%%%%%%%%%%%%%%%%%%%%%%%%%%%%%%%%%%%%%%%%%%%%%%%%%%%%%%%%%%%%

\section{Conclusions and Outlook}\label{sec:conclusion}

We have proposed a framework for segmentation-based compression 
that can outperform existing codecs for piecwiese smooth data by a 
considerable margin. It is remarkable that this can be achieved 
with very simple concepts, provided that they are chosen carefully. 
For a given density and quantisation level, the full method only 
requires a single, intuitive parameter: the segment cost weight
$\lambda$. This makes it easy to use.  
From a more general viewpoint, the success of our approach can serve as
one more example which shows that transparent energy-based modelling 
should be preferred over ad hoc algorithms.

\medskip
Our penaliser can be interpreted as a coding cost term for the segment 
boundaries. In our ongoing work, we are extending this
idea to a full energy-based rate distortion framework that also 
optimises the additional coding costs of the inpainting data. 
Moreover, we are also applying it to the compression of optic 
flow fields for video coding.

%%%%%%%%%%%%%%%%%%%%%%%%%%%%%%%%%%%%%%%%%%%%%%%%%%%%%%%%%%%%%%%%%%%%%%%%%%%

\bibliographystyle{IEEEtran}
\bibliography{myrefs}

\end{document}